\documentclass[aps,prl,twocolumn,superscriptaddress,showpacs,showkeys,10pt,longbibliography]{revtex4-1}
\usepackage{mathrsfs}
\usepackage{graphicx}
\usepackage{amsmath}
\usepackage{xcolor}
\usepackage{makecell}
\usepackage{multirow}
\usepackage{booktabs} 
\usepackage{tabstackengine}

\newcommand{\refeq}[1]{Eq.~(\ref{#1})}
\newcommand{\reffig}[1]{Fig.~\ref{#1}}

\newcommand\scalemath[2]{\scalebox{#1}{\mbox{\ensuremath{\displaystyle #2}}}}

\usepackage[colorlinks]{hyperref}
\hypersetup{
	plainpages=true,
	breaklinks=true,
	hypertexnames=false,
	pageanchor=true,
	colorlinks=true,
	linkcolor={blue},
	citecolor={blue},
	urlcolor={blue},
	anchorcolor={black}
}

\begin{document}
\bibliographystyle{unsrt}
\title{Efficient quantum state tomography with auxiliary Hilbert space}
\author{Ruifeng Liu}
\thanks{These two authors contributed equally}
\affiliation{Shaanxi Province Key Laboratory of Quantum Information and Quantum Optpelectronic Devices, School of Science, Xi'an Jiaotong University, Xi'an 710049, P.R. China}
\author{Junling Long}
\thanks{These two authors contributed equally}
\affiliation{National Institute of Standards and Technology, Boulder, Colorado 80305, USA}
\affiliation{Department of Physics, University of Colorado, Boulder, Colorado 80309, USA}

\author{Pei Zhang}\email{zhang.pei@xjtu.edu.cn}
\affiliation{Shaanxi Province Key Laboratory of Quantum Information and Quantum Optpelectronic Devices, School of Science, Xi'an Jiaotong University, Xi'an 710049, P.R. China}

\author{Russell E. Lake}
\affiliation{Bluefors Oy, Arinatie 10, 00370 Helsinki, Finland}

\author{Hong Gao}
\affiliation{Shaanxi Province Key Laboratory of Quantum Information and Quantum Optpelectronic Devices, School of Science, Xi'an Jiaotong University, Xi'an 710049, P.R. China}

\author{David P. Pappas}
\affiliation{National Institute of Standards and Technology, Boulder, Colorado 80305, USA}
\affiliation{Department of Physics, University of Colorado, Boulder, Colorado 80309, USA}
\author{Fuli Li}\email{flli@xjtu.edu.cn}
\affiliation{Shaanxi Province Key Laboratory of Quantum Information and Quantum Optpelectronic Devices, School of Science, Xi'an Jiaotong University, Xi'an 710049, P.R. China}

\begin{abstract}
Quantum state tomography is an important tool for quantum communication, computation, metrology, and simulation. Efficient quantum state tomography on a high dimensional quantum system is still a challenging problem. Here, we propose a novel quantum state tomography method, auxiliary Hilbert space tomography, to avoid pre-rotations before measurement in a quantum state tomography experiment. Our method requires projective measurements in a higher dimensional space that contains the subspace that includes the prepared states. We experimentally demonstrate this method with orbital angular momentum states of photons. In our experiment, the quantum state tomography measurements are as simple as taking a photograph with a camera. We experimentally verify our method with near-pure- and mixed-states of orbital angular momentum with dimension up to $d=13$, and achieve greater than 95~\% state fidelities for all states tested. This method dramatically reduces the complexity of measurements for full quantum state tomography, and shows potential application in various high dimensional quantum information tasks. 
\end{abstract}

\maketitle


Quantum information science promises new and extraordinary types of communication and computation due to the unique properties of quantum superposition and entanglement. Characterization and validation of a synthetic quantum system requires experimental measurements of the quantum state that is described by its density matrix. Quantum state tomography (QST) reconstructs the density matrix of a quantum system by measuring a full set of observables that span the entire state space of the system~\cite{Paris04,siah2015introduction}. To perform full QST on a system with dimension $d$, the required number of unitary rotations combined with projection measurements is $d^2-1$. As a result, the resources required for brute-froce QST scales unfavorably with the dimension size, making QST unfeasible for high dimensional quantum systems~\cite{haffner2005scalable,stefano2017determination,ansari2018tomography,titchener2018scalable}. 

The past years have witnessed considerable effort devoted to boosting the efficiency of QST~\cite{Okamoto:2012aa,smolin2012efficient,Mahler:2013aa,Hou_2016,martinez2019experimental} or to developing alternative methods of characterizing a quantum system. Various methods of avoiding full QST have been developed to address cases where only partial knowledge of the density operator is sufficient to capture the key aspects of a system. These approximations include entanglement witness~\cite{Barbieri:2003aa,Bourennane:2004aa}, purity estimation~\cite{bagan2005purity}, fidelity estimation~\cite{flammia2011direct}, and quantum-state classification~\cite{lu2018separability,gao2018experimental}. In addition, more manageable QST protocols that require measurement of fewer observables have been studied for states possessing special properties such as permutationally invariant QST for permutationally invariant states~\cite{toth2010permutationally}, and compressed sensing~\cite{Gross:2010aa,Liu:2012aa,ahn2019adaptive}. 

\begin{figure*}[htb]
\centerline{\includegraphics[width=16cm]{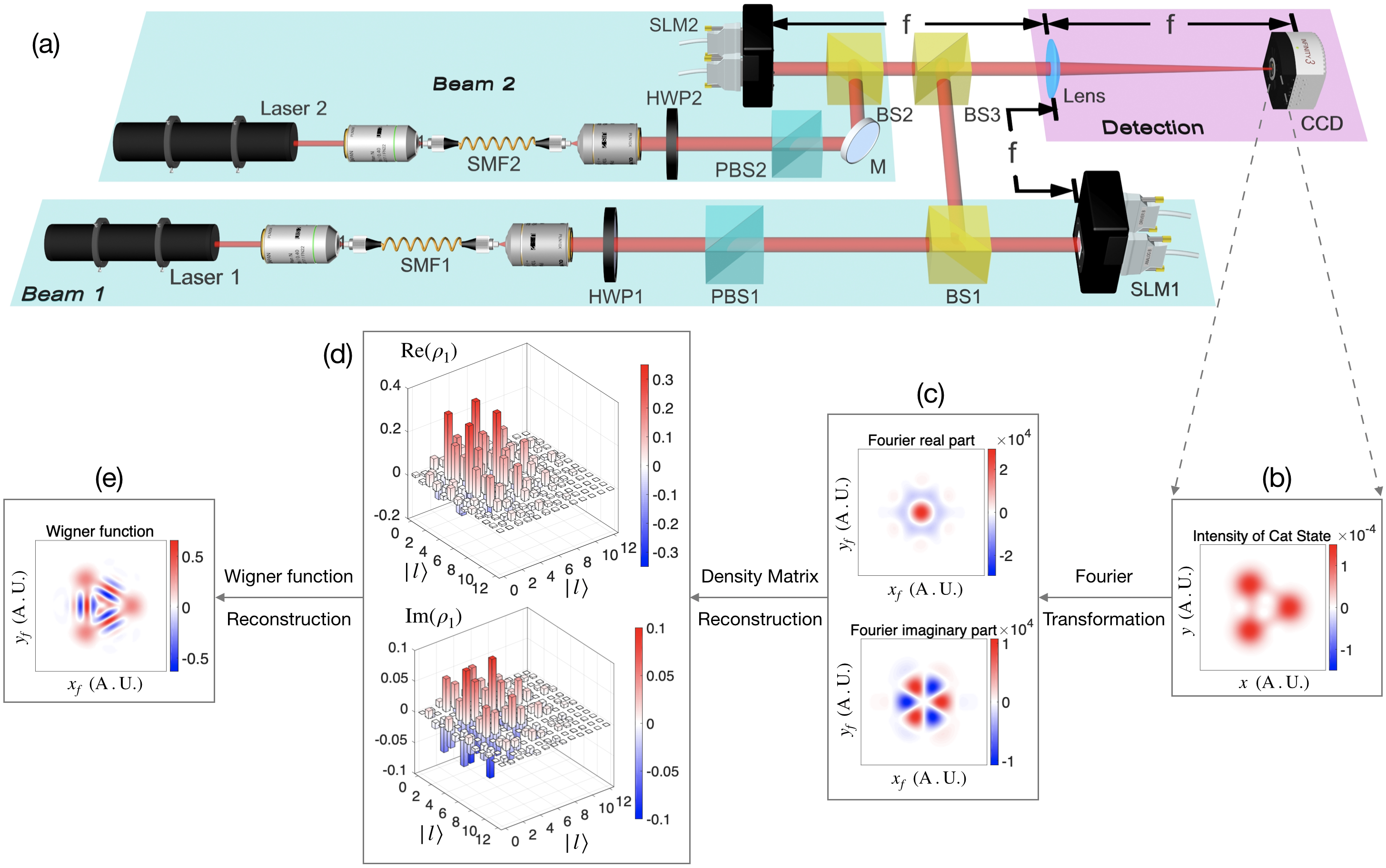}}
\caption{(a) Experimental setup for generation and tomography of OAM states. Two beams from independent lasers are individually cleaned and collimated with single mode fibers (SMFs) and objectives. Transmitted power is controlled with a combination of half wave plates (HWPs) and polarization beam splitters (PBSs). Two superposition OAM modes are individually generated by two phase-only spatial light modulators (SLMs) with computer-generated holograms (CGHs)~\cite{bolduc2013exact}. The two beams are then combined in beam splitter-3 (BS3). A charge coupled device (CCD) is used to record the intensity profile of the first-diffracted order field from the SLMs. We implement a far field of the generated beam by a lens. (b) The intensity distribution of a simulated three-component cat state in real space~\cite{gribbin2011search, vlastakis2013deterministically, liu2019classical}. (c) The real and imaginary parts of the Fourier transformation of the intensity pattern in (b). (d) and (e) are the reconstructed density matrix and Wigner function, respectively.}
\label{setup}
\end{figure*}

In this article we propose a new method to reconstruct the full density matrix that requires no rotation before the projection measurement. We refer to our method as auxiliary Hilbert space tomography (AHST). Suppose an unknown state $\hat{\rho}$ is prepared in a $d$-dimensional Hillbert space, $\mathcal{H}_{d}$, spanned by orthonormalized sates $|\phi_i\rangle$, where $i=1, 2, \cdots, d$. Outcomes of measurements performed on copies of the state $\hat{\rho}$ are given by the Born rule $P(i)=\mathrm{tr}(\hat{\rho}\Pi_{i})$, where $\{\Pi_{i}\}$ is a positive operator-valued measure (POVM) that describes the measurement setup. If the statistics of the outcome probabilities uniquely determine the state $\hat{\rho}$, the POVM is informationally complete (IC). It is obvious that the minimum number of rank-1 elements of an IC-POVM for $\hat{\rho}$ is $d^2$. However, in most cases, a directly experimental accessible POVM consists of projective measurements $|\phi_i\rangle\langle\phi_i|$, which has in total $d$ elements. As a result, rotations are needed to access other elements to form an IC-POVM. In our AHST protocol, instead of performing projective measurements in the original Hilbert space, $\mathcal{H}_{d}$, we conduct the projective measurements in another higher dimensional Hilbert space, $\mathcal{H}_{D}$, that contains $\mathcal{H}_{d}$ as a subspace and is spanned by basis $|\psi_j\rangle$, where $j=1, 2, \cdots, D$. If the projective elements $\{ \mathcal{I}_{d}|\psi_j\rangle\langle\psi_j|\mathcal{I}_{d}\}$, where $\mathcal{I}_{d}=\sum_{i=1}^{d}|\phi_i\rangle\langle\phi_i|$, forms an IC-POVM in $\mathcal{H}_{d}$, no rotations would be required to reconstruct the density matrix $\hat{\rho}$ in $\mathcal{H}_{d}$. Thus, the dimension size of the measurement space $\mathcal{H}_{D}$ needs to satisfy $D\geq d^2$.

Here, we apply the AHST protocol to optical system to demonstrate full QST of the orbital angular momentum (OAM) states of photons. The OAM of photons was discovered in 1992 on a family of paraxial light beams~\cite{allen1992orbital}. These paraxial light beams have helical phase front of $e^{il\varphi}$ and carry well-defined OAM of $l\hbar$ per photon, where $l$ can take any integer value. Extensive studies~\cite{Erhard_2017} in recent years have proposed using OAM states for a range of exciting applications in quantum information~\cite{Nagali_2009}, and free-space optical communications~\cite{Gibson_2004,lavery2017free}. For example, researchers have found that OAM states promise higher information capacity per-particle~\cite{Leach:2012aa}, more efficient quantum circuit structure~\cite{Bocharov:2017aa}, enhanced robustness against eavesdropping and quantum cloning~\cite{Bechmann-Pasquinucci:2000aa}, complex entanglement structures~\cite{Huber:2013aa,Roslund_2013,Malik_2016}, and robust resilience to noise and loss~\cite{Cerf:2002aa,Vertesi:2010aa}. Meanwhile, tomography of the high dimensional OAM states have also been studied extensively with POVMs~\cite{nicolas2015quantum}, and mutually unbiased bases~\cite{Giovannini:2013aa,Bent:2015aa}.

Typical laser modes that carry OAM are Laguerre-Gaussian (LG) modes, labelled by two mode indices, $p$ and $l$. The modal index $l$ determines the OAM carried by each photon while $p$ defines the radial intensity distribution on the cross section of this mode and only takes non-negative integer values. A photon that is in an LG eigenmode can be represented as $|p,l\rangle$. According to the AHST protocol, we can prepare an OAM state in the space spanned by $|p,l\rangle$ basis and measure it using the polar coordinate basis $|r,\phi\rangle$, i.e., recording the intensity with a charge-coupled device (CCD) camera on the beam cross section. However, both $|p,l\rangle$ and $|r,\phi\rangle$ bases have infinite dimensions, and they are equivalent as complete orthogonal bases for paraxial beams. To ensure that the prepared states are in a subspace of the measurement space, we restrict the prepared states in basis specified by,
\begin{equation}       
p=0,\ l\geq0
\label{subspace}               
\end{equation}
For the rest of the article, we only focus on OAM states that satisfy \refeq{subspace}, and use $|l\rangle$ to represent the state $|p=0,l\geq0\rangle$.

The intensity measurement results of an arbitrary state $\hat{\rho}$ prepared in the basis $|l\rangle$ by using a CCD camera at beam waist can be described by,
\begin{equation}
\begin{split}
I(r,\phi)&=A \langle r,\phi|\hat{\rho}| r,\phi\rangle,\\
&=A \sum_{l_{1}=0}^{\infty}\sum_{l_{2}=0}^{\infty} \Psi_{l_{1}}(r,\phi)\Psi_{l_{2}}^{*}(r,\phi)\hat{\rho}_{l_{1},l_{2}}
\end{split}
\label{Inten_OAM}               
\end{equation}
where $A$ is a scale factor that can be determined by normalization of the density matrix, $\hat{\rho}_{l_{1},l_{2}}$ is the density matrix element in $|l\rangle$ basis, and $\Psi_{l}(r,\phi)$ is the complex amplitude of the laser mode that is in state $|l\rangle$ (see Method for detailed expression). Then, we apply a linear (Fourier) transformation to both side of \refeq{Inten_OAM} \cite{prabhakar2011revealing} and get,
\begin{equation}
\begin{split}
\mathscr{F}[I(r,\phi)]&=A \sum_{l_{1}=0}^{\infty}\sum_{l_{2}=0}^{\infty} \mathscr{F}[\Psi_{l_{1}}(r,\phi)\Psi_{l_{2}}^{*}(r,\phi)]\hat{\rho}_{l_{1},l_{2}}\\
&=A \sum_{l_{1}=0}^{\infty}\sum_{l_{2}=0}^{\infty} P_{l_{1},l_{2}}(r_f,\phi_f)\hat{\rho}_{l_{1},l_{2}}
\end{split}
\label{Fourier_I}               
\end{equation}
where $\mathscr{F}[\cdot]$ is the Fourier transform, $r_f$ and $\phi_f$ are the polar coordinates in the Fourier plane, and $P_{l_{1},l_{2}}(r_f,\phi_f) = \mathscr{F}[\Psi_{l_{1}}(r,\phi)\Psi_{l_{2}}^{*}(r,\phi)]$. We find that $P_{l_{1},l_{2}}(r_f,\phi_f)$ function has the following orthogonal property,
\begin{equation}
\begin{split}
&\int_{0}^{\infty}\int_{0}^{2\pi} P_{l_{1},l_{2}}(r_f,\phi_f)P_{l_{1}^{'},l_{2}^{'}}^{*}(r_f,\phi_f)e^{\frac{\pi^2r_f^2\sigma^2}{2}}C_{l_{1}^{'},l_{2}^{'}}r_f\text{d}r_f\text{d}\phi_f\\
&= \delta_{l_{1},l_{1}^{'}}\delta_{l_{2},l_{2}^{'}}
\end{split}
\label{Orthogonal_OAM}               
\end{equation}
where $C_{l_{1}^{'},l_{2}^{'}}$ is a normalization constant determined by $l_{1}^{'}$ and $l_{2}^{'}$; $\sigma$ is the beam waist which can be measured experimentally (see Methods). Given \refeq{Orthogonal_OAM}, the density matrix can be calculated in the following way,
\begin{equation}
\begin{split}
\hat{\rho}_{l_{1},l_{2}} = &\frac{C_{l_{1},l_{2}}}{A}\int_{0}^{\infty}\int_{0}^{2\pi} \mathscr{F}[I(r,\phi)]P_{l_{1},l_{2}}(r_f,\phi_f)\\
&\times e^{\frac{\pi^2r_f^2\sigma^2}{2}}r_f\text{d}r_f\text{d}\phi_f
\end{split}
\label{GetDensityMatrix}               
\end{equation}
The above equation proves that, in theory when $r$ and $\phi$ are continuous, projective measurements $\{|r,\phi\rangle\langle r,\phi|\}$ in coordinate space indeed form an IC-POVM in the OAM subspace subject to \refeq{subspace}. \reffig{setup}(b)-(e) show the flow diagram of AHST protocol. We simulated the intensity of a three-component OAM cat state [\reffig{setup}(b)], $|\alpha\rangle+e^{i0.6\pi}|e^{i2\pi/3}\alpha\rangle+e^{-i0.3\pi}|e^{i4\pi/3}\alpha\rangle$~\cite{gribbin2011search, vlastakis2013deterministically, liu2019classical}, where $|\alpha\rangle$ is an OAM coherent state defined by $|\alpha\rangle=e^{-\frac{|\alpha|^2}{2}}\sum_{l=0}^{\infty}\frac{\alpha^{l}}{\sqrt{l!}}|l\rangle$, and $\alpha=2$. After applying the Fourier transformation on the intensity pattern we observe its complex amplitude as shown in \reffig{setup}(c). Using \refeq{GetDensityMatrix}, the density matrix can be reconstructed as shown in \reffig{setup}(d). \reffig{setup}(e) shows the Wigner function calculated from the reconstructed density matrix.

\begin{figure*}[htb]
\centerline{\includegraphics[width=16cm]{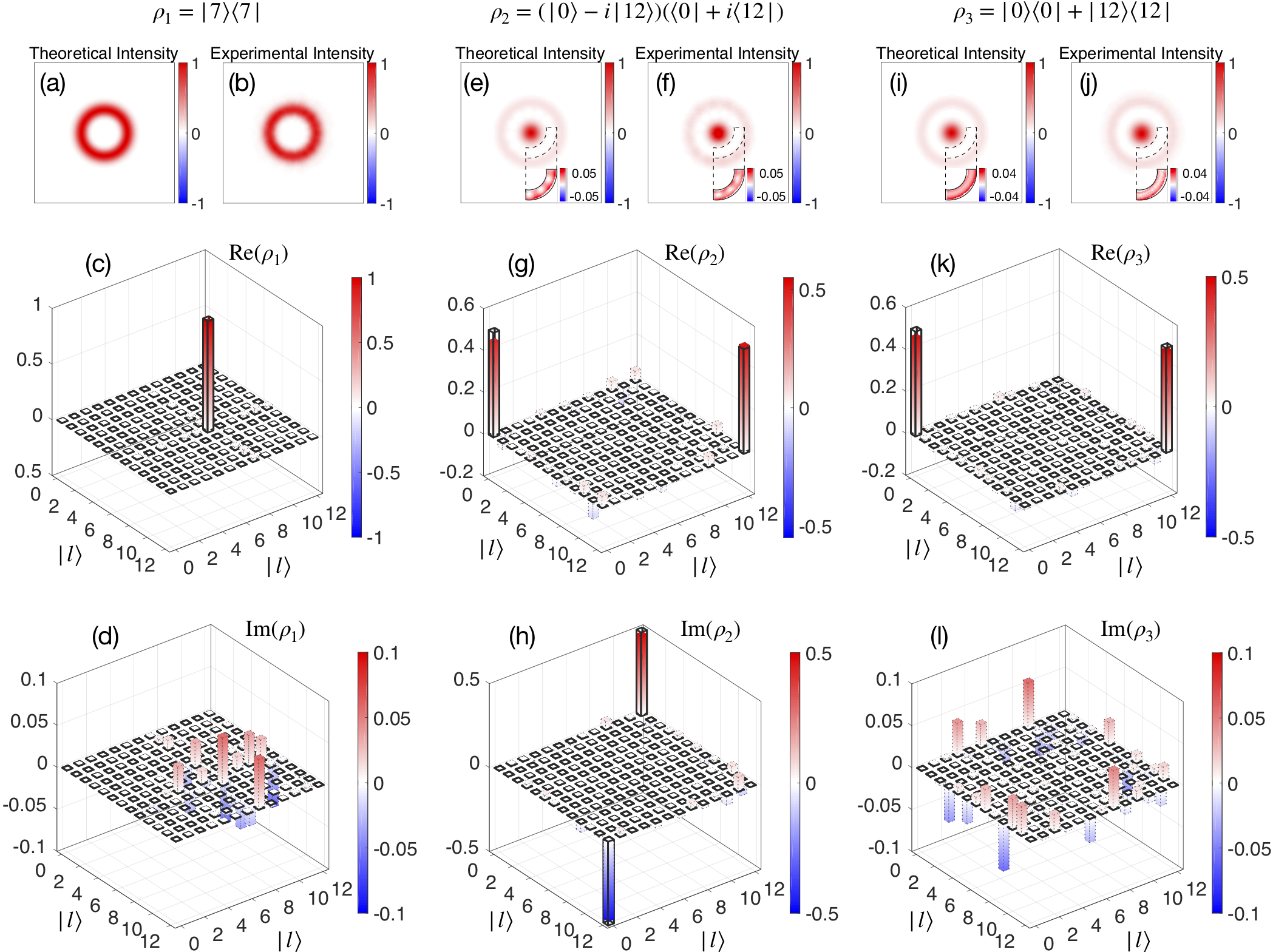}}
\caption{AHST of pure OAM eigenstates, a superposition state, and a mixed state.  (a), (e), and (i) are the numerically simulated intensity patterns for states $\rho_1=|7\rangle\langle7|$, $\rho_2=(|0\rangle-i|12\rangle)(\langle0|+i\langle12|)$ and $\rho_3=|0\rangle\langle0|+|12\rangle\langle12|$, respectively. (b), (f), and (j) are the experimentally recorded intensity patterns for three different states. The insets show the zoomed-in intensity distributions of different states. Axis labels in (a), (b), (e), (f), (i), and (j) follow the axis labels in \reffig{setup} (b). Note that the horizontal and vertical axes in (a) and (b) have a different scale than those in (e), (f), (i), and (j) for better appearance. (c), (g), and (k) represent the real parts of the reconstructed density matrix. (d), (h), and (l) represent the imaginary parts of the density matrix. The theoretical values and experimental values are shown by empty bars with solid line borders and color bars with dashed line borders, respectively. Reconstructed density matrices are expanded in OAM eigenstates $|l\rangle$ ranges from $\{|0\rangle, \cdots, |12\rangle\}$.}
\label{data1}
\end{figure*}

We experimentally generated various OAM states to test our AHST protocol. \reffig{setup}(a) shows a schematic of the experimental setup for generating and detecting OAM states subject to \refeq{subspace} (see Methods for more details). In a first experiment, we generate different OAM eigenstates $|l\rangle$ with laser 1 turned on and laser 2 off. Density matrices are obtained by performing AHST using \refeq{GetDensityMatrix} and we choose the first thirteen OAM eigenstates, $|l=0, 1, \cdots, 12\rangle$ ,to span our tomography space. However, \refeq{GetDensityMatrix} does not guarantee a physical density matrix, i.e., a Hermitian positive semi-definite density matrix. To remedy this, we use the least squares method to find the physical density matrix that is closest to the unphysical experimentally-obtained density matrix (details in Methods). 
\reffig{data1}(a) - (d) demonstrate the predicted intensity, measured intensity, real and imaginary part of the density matrix of one OAM eigenstate $\left|7 \rangle\langle7\right|$, respectively. We calculate the state fidelity using $F=(\mathrm{Tr}\sqrt{\sqrt{\hat{\rho}^{t}}\hat{\rho}^{e}\sqrt{\hat{\rho}^{t}}})^2$, where $\hat{\rho}^{\mathrm{e}}$ and $\hat{\rho}^{\mathrm{t}}$ are the experimentally reconstructed and theoretical density matrices, respectively. For the state $\left|7 \rangle\langle7\right|$, we extracted a state fidelity $0.980\pm0.003$. Table I summarizes the state fidelities for some other eigenstates, and all of them show relatively high state fidelities.

In a second experiment, we test our AHST with OAM superposition states. \refeq{superposition} describes the three superposition states generated in the experiment,
\begin{equation}
\begin{split}
&|\psi _{\mathrm{G}}\rangle=\frac{1}{\sqrt{2}}(|0\rangle-i|12\rangle)\\
&|\psi_{\mathrm{c}}\rangle =C_{c}\sum_{l=0}^{6}\frac{\alpha^{2l}}{\sqrt{(2l)!}}|2l\rangle\\
&|\psi _{\mathrm{s}}\rangle=C_{s}\sum_{l=0}^{6}(\mathrm{-tanh}\gamma)^{l}\frac{\sqrt{(2l)!}}{2^l l!}|2l\rangle\\
\end{split}
\label{superposition}
\end{equation}
\noindent where $C_{c}$ and $C_{s}$ are the normalization factors of the states and $\alpha$, $\gamma$ are the parameters of the states.

\begin{figure*}[htb]
\centerline{\includegraphics[width=16cm]{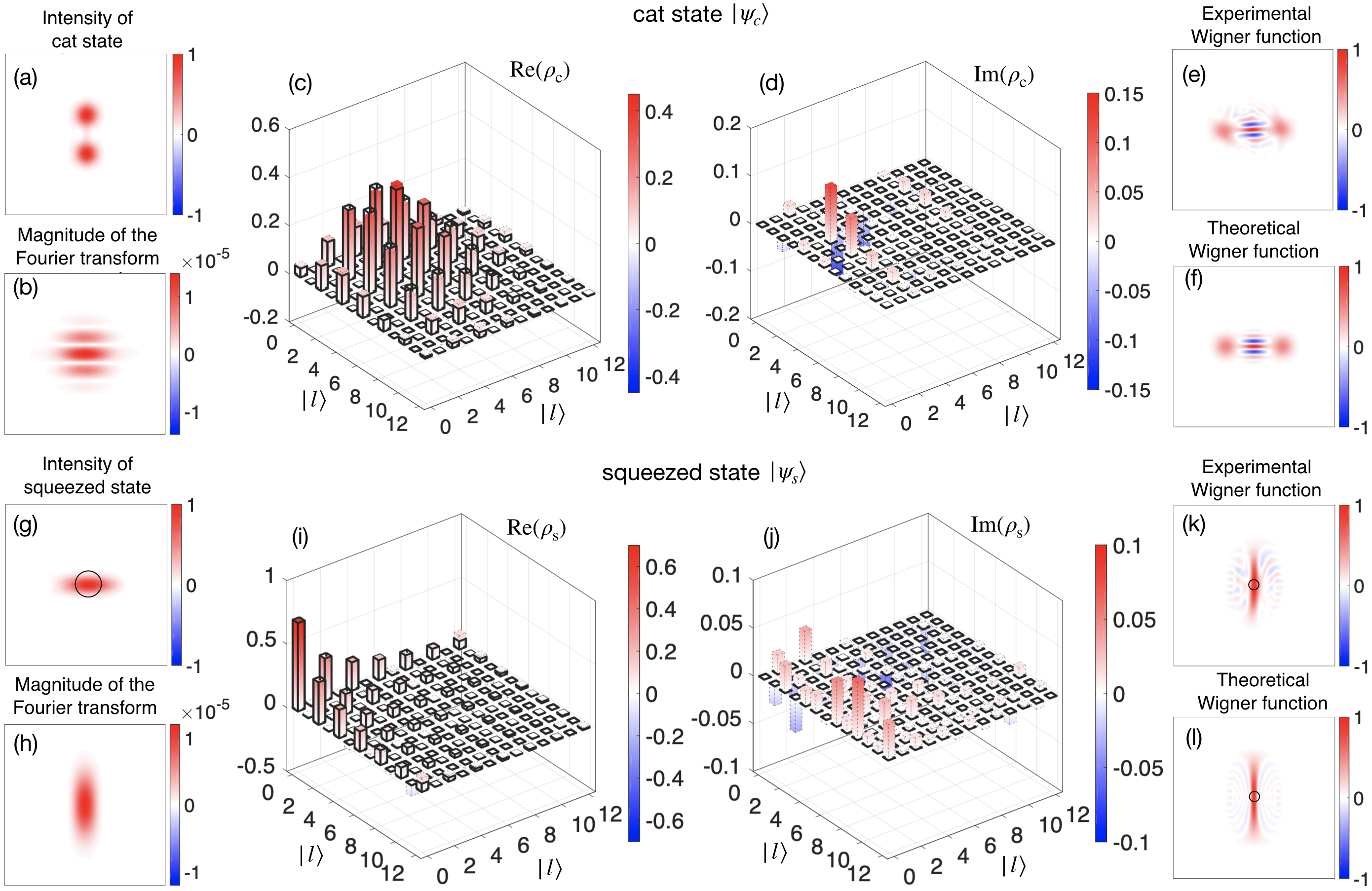}}
\caption{AHST of the cat state, $|\psi\rangle_c$, and the squeezed state, $|\psi\rangle_s$, defined in \refeq{superposition}. (a) and (g) are the intensity patterns of $|\psi\rangle_c$ and $|\psi\rangle_s$, whose axis labels follow the axis labels in \reffig{setup} (b). (b) and (h) are the magnitude of the Fourier transforms of the intensity patterns, whose axis labels follow the axis labels in \reffig{setup} (c). (c) and (i) represent the real parts of reconstructed density matrices for $|\psi\rangle_c$ and $|\psi\rangle_s$. (d) and (j) represent the imaginary parts of reconstructed density matrices. (e) and (k) are the experimentally reconstructed Wigner functions. (f) and (l) are the theoretical Wigner functions. Axis labels in (e), (f), (k), and (l) follow the axis labels in \reffig{setup} (e). The faint patterns of ripples extending from the origin in (k) and (l) are caused by truncation at $l=12$ of the density matrix used to represent the squeezed state. The black solid circles in (g), (k) and (l) show the full-width at half-maximum of the Gaussian mode (OAM vacuum state, $|l=0 \rangle$) in corresponding representations.}
\label{data2}
\end{figure*}

One may observe that the definitions of $|\psi\rangle _{\mathrm{c}}$ and $|\psi\rangle _{\mathrm{s}}$ are the same as truncated cat state~\cite{gribbin2011search} and squeezed state~\cite{gribbin2011search} of a one-dimensional harmonic oscillator, respectively. We will discuss the relation between these OAM superposition states and the non-classical states of a one-dimensional harmonic oscillator in the discussion section. \reffig{data1}(e) - (h) demonstrate the tomography results of the superposition state $|\psi _{\mathrm{G}}\rangle \langle\psi _{\mathrm{G}}|$. The fidelity of this superposition state is $0.961\pm 0.004$. This indicates that our system can generate high quality pure states and validates the AHST protocol. \reffig{data2}(a)-(d) and (g)-(j) shows the experimental results for a cat state with $\alpha=2$ and a squeezed state with $\gamma=1.5$.

Finally, we perform full QST on different mixed states using the AHST protocol. As shown in Fig~(\ref{setup}), to generate the mixed states, we turn on both independent He-Ne lasers. Each mixed state is a classical mixture of two near-pure states that are generated by the spatial light modulators (SLMs) in each arm. By rotating the half wave plates (HWPs) in each arm, we can adjust the weight of the two superposition states in a mixed state. In our experiment, we generate the two mixed states,
\begin{equation}
\begin{split}
&{ \hat{\rho}  }_{ \mathrm{m1 }}=\frac{1}{2}\left|0 \rangle\langle 0 \right|+\frac{1}{2}\left|12 \rangle\langle 12\right|\\
&{ \hat{\rho}  }_{ \mathrm{m2 }}= \frac{1}{4}(|0\rangle+e^{i\frac{4\pi}{3}}|12\rangle)(\langle0|+e^{-i\frac{4\pi}{3}}\langle12|)+  \frac{1}{2}|6\rangle\langle6|
\end{split}
\label{mixed}
\end{equation}

The experimental results of the state ${\hat{\rho}  }_{ \mathrm{m1 }}$ are shown in \reffig{data1}(i) - (l) (see Table.~\ref{tab:1} for the result of the mixed state ${\hat{\rho}  }_{ \mathrm{m2 }}$). The fidelity of this mixed state is $0.955\pm 0.004$. It is interesting to compare the states $|\psi _{\mathrm{G}}\rangle \langle\psi _{\mathrm{G}}|$ and ${\hat{\rho}  }_{ \mathrm{m1 }}$ because both of them just consist of OAM eigenstates $|0 \rangle$ and $|12 \rangle$. The intensity patterns of these two states in \reffig{data1}(e), (f), (i), and (j) look extremely similar to each other. However, when we zoom into the white-ring parts [the insets in \reffig{data1}(e), (f), (i), and (j)], we find that there are very small azimuthal interference patterns for the superposition state $|\psi _{\mathrm{G}}\rangle \langle\psi _{\mathrm{G}}|$, which captures the coherence between $|0 \rangle$ and $|12 \rangle$ state components. This small coherent signal is resolved by our AHST tomography protocol. This is verified by the non-zero off-diagonal terms in the imaginary part of the density matrix shown in  \reffig{data1}(h). The physics behind this interference feature is very similar to the holography technique. In fact, both holography technique and the AHST protocol encode information as interference patterns. However, the interference pattern in a hologram comes from a target light field and a known reference light field, whereas in the AHST protocol, there is no reference light field and the interference is between all the different OAM eigenmodes contained in the state to be measured. In other words, the AHST protocol is self-referenced. 

Table.~\ref{tab:1} is a summary of the fidelities of the reconstructed density matrices for all of the different OAM states tested in our experiments. In these examples, all fidelities are greater than $95~\%$. As shown in Table~\ref{tab:1}, eigenstates with higher OAM number show lower fidelities due to the fact that higher order OAM modes are more sensitive to the astigmation of the lens and and the flatness of the SLMs. For mixed states, the main error sources are intensity fluctuations of the two lasers and the overlap inaccuracy of the two independent beams.

\begin{table}[htbp]
\centering
\caption{\label{tab:1}Experimental fidelities of the reconstructed density matrices, the uncertainties are characterized by the standard deviation of 10 repetitions.}
\setlength{\tabcolsep}{6mm}{
\begin{tabular}{ccc}
\toprule[1pt]
 \multirow{2}{*}{\ \ }&\multirow{2}{*}{States}&\multirow{2}{*}{Fidelity} \\
 &&\\
\midrule[1pt]
 \multirow{13}{*}{Eigenstate} &  \multirow{13}{*}{\makecell[c]{$\left| 0 \right>$ \\ $\left| 1 \right>$ \\ $\left| 2 \right>$ \\ $\left| 3 \right>$ \\ $\left| 4 \right>$ \\ $\left| 5 \right>$ \\ $\left| 6 \right>$ \\ $\left| 7 \right>$ \\ $\left| 8 \right>$ \\ $\left| 9 \right>$\\ $\left| 10 \right>$\\ $\left| 11 \right>$\\ $\left| 12 \right>$}}&  \multirow{13}{*}{\makecell[c]{$0.996\pm0.001$ \\ $0.995\pm0.001$ \\ $0.992\pm0.001$ \\ $0.991\pm0.002$ \\ $0.988\pm0.001$ \\ $0.985\pm0.002$ \\ $0.982\pm0.002$ \\ $0.980\pm0.003$ \\ $0.977\pm0.003$ \\ $0.974\pm0.004$\\ $0.967\pm0.005$\\ $0.959\pm0.004$\\ $0.953\pm0.005$}} \\
 &&\\&&\\ &&\\&&\\ &&\\&&\\ &&\\&&\\&&\\&&\\&&\\&&\\
\midrule[0.5pt]
 \multirow{3}{*}{Superposition} &  \multirow{3}{*}{\makecell[c]{$|\psi\rangle _{\mathrm{G}}$ \\ $|\psi\rangle _{\mathrm{c}}$ \\$|\psi\rangle _{\mathrm{s}}$}}&  \multirow{3}{*}{\makecell[c]{$0.961\pm0.004$ \\ $0.969\pm0.002$\\ $0.975\pm0.002$}} \\
 &&\\&&\\
 \midrule[0.5pt]
 \multirow{2}{*}{Mixed State} &  \multirow{2}{*}{\makecell[c]{$\rho_{\mathrm{m1}}$ \\ $\rho_{\mathrm{m2}}$}}&  \multirow{2}{*}{\makecell[c]{$0.955\pm0.004$ \\ $0.952\pm0.005$}} \\
 &&\\
\bottomrule[1pt]
\end{tabular}}
\end{table}
\section{Discussion}
It is known that LG modes are eigenmodes of a two-dimensional isotropic harmonic oscillator. The subspace formed by LG modes subject to \refeq{subspace} looks very similar to Hilbert space formed by Fock states $|n\geq0\rangle$, which are eigenstates of a one-dimensional harmonic oscillator. In fact, we can make a bijective mapping between states $|l\rangle$ studied here and Fock states $|n\rangle$ by setting $l=n$. We will show that there exists a close relation between the Fourier transform of the intensity [\refeq{Fourier_I}] and the Wigner function~\cite{scully1999quantum} of Fock states, which was originally introduced to provide a phase-space representation of a quantum state in a continuous Hilbert space. To prove this, we substitute the detailed formula of function $P_{l_{1},l_{2}}(r_f,\phi_f)$ (see Methods) into \refeq{Fourier_I} and get,
\begin{equation}
\begin{split}
&\mathscr{F}[I(r,\phi)]=A\sum_{l_{1}=0}^{\infty}\sum_{l_{2}=0}^{\infty}\hat{\rho}_{l_{1},l_{2}} (-i)^{|l_{1}-l_{2}|} \frac{\sqrt{l_{1}!l_{2}!}}{\mathrm{max}(l_{1},l_{2})!}\\
&\times e^{-2R^{2}} (\sqrt{2} R)^{|l_{1}-l_{2}|}L_{\mathrm{max}(l_{1},l_{2})}^{|l_{1}-l_{2}|}(2 R^{2})e^{i(l_{1}-l_{2})\phi_f},
\end{split}
\label{Fourier_I_detailed}               
\end{equation}
where $L_{\mu}^{\upsilon}(\cdot)$ is the associated Laguerre function and $R=\frac{\pi\sigma r_f}{2}$. The Wigner function of an arbitrary one-dimensional harmonic oscillator state $\hat{\rho}'$ in Fock state basis is,
\begin{equation}
\begin{split}
&W(p,q)=A'\sum_{n_{1}=0}^{\infty}\sum_{n_{2}=0}^{\infty}\hat{\rho}'_{n_{1},n_{2}} (-1)^{\mathrm{min}(n_{1},n_{2})} \frac{\sqrt{n_{1}!n_{2}!}}{\mathrm{max}(n_{1},n_{2})!}\\
&\times e^{-r'^{2}} (\sqrt{2} r')^{|n_{1}-n_{2}|}L_{\mathrm{min}(n_{1},n_{2})}^{|n_{1}-n_{2}|}(2 r'^2)e^{i(n_{1}-n_{2})\phi '}.
\end{split}
\label{Wigner}               
\end{equation}
In this expression $A'$ is a normalization factor, $n_{1}$ and $n_{2}$ are the indices of Fock states $|n_1\rangle$ and $|n_2\rangle$, $p$ and $q$ are the two quadratures in the phase space, and $r'$ and $\phi '$ are given by  $r '=\sqrt{\frac{1}{\hbar}(p^{2}+q^{2})}$ and $\phi '=\mathrm{arctan}(\frac{p}{q})$. 

We can see that \refeq{Fourier_I_detailed} and \refeq{Wigner} share a very similar form except for the constant factors, $(-i)^{|l_{1}-l_{2}|}$ versus $(-1)^{\mathrm{min}(n_{1},n_{2})}$, and the Gaussian terms, $e^{-2 R^{2}}$ versus $e^{-r'^{2}}$. The difference in the Gaussian terms can always be reconciled by dividing both sides of \refeq{Fourier_I_detailed} by a known term $e^{-R^{2}}$. However, the constant factor $(-i)^{|l_{1}-l_{2}|}$ makes $\mathscr{F}[I(r,\phi)]$ a complex function compared to the real-valued Wigner function. Although both $\mathscr{F}[I(r,\phi)]$ and $W(p,q)$ uniquely determine the state they represent, the real-valued Wigner function $W(p,q)$ is a more standard representation. Fortunately, we can still use \refeq{Wigner} to construct the same Wigner function for OAM states studied here given the density matrix reconstructed from \refeq{Fourier_I_detailed}.

\reffig{data2}(a) - (l) show the intensity, $|\mathscr{F}[I(r,\phi)]|$, the density matrix, and the Wigner function $W(p,q)$ of the states $|\psi _{\mathrm{c}}\rangle$ and $|\psi_{\mathrm{s}}\rangle $, respectively. The plots of their Wigner functions indeed indicate that they are OAM cat and squeezed states. However, we note that the ``OAM vacuum state'' is not a real vacuum state but is actually the fundamental Gaussian mode of the laser. Thus, the OAM cat state is a combination of two displaced Gaussian laser modes and the squeezed state is only squeezed with respect to the Gaussian mode laser spot [black solid circles in \reffig{data2}(g), (k) and (l)]. However, compared to the Fock cat and squeezed states, the OAM cat and squeezed states are more straightforward to generate. In addition, our AHST protocol also makes the QST measurement as easy as taking a photo. With the ease in generation and tomography, the OAM states subject to \refeq{subspace} provide a great platform for simulating quantum optics phenomenons related to Fock states.

We noticed that, In reference~\cite{liu2019classical}, the authors also utilized the OAM states to generate a macroscopic Schr$\ddot{\mathrm{o}}$dinger cat. However, the experimental complexity of state tomography using AHST is greatly reduced compared to the method of mutually unbiased measurements that is used in reference~\cite{liu2019classical}.

In conclusion, we proposed an efficient full QST protocol, AHST, and experimentally demonstrated it with the OAM states of photons. The tomography results show very high state fidelities with both near-pure and mixed states. We also observe a surprising similarity between OAM states in the subspace required by AHST protocol and Fock states. By exploiting this similarity, we generated OAM cat and squeezed states and constructed their Wigner functions. The AHST protocol may be extended to other quantum systems to simplify the QST in general. Our results for OAM systems may also lead to interesting research on simulating quantum optics phenomena with OAM states.

\section{Acknowledgements}
J.L. acknowledges discussions with E. H. Knill, S. C. Glancy, Z. Dutton, and HS Ku. This work is supported by the National Natural Science Foundation of China (NSFC) (11534008, 11605126, 91736104); Ministry of Science and Technology of the Peoples Republic of China (MOST) (2016YFA0301404); Natural Science Basic Research Plan in Shaanxi Province of China (2017JQ1025); Doctoral Fund of Ministry of Education of China (2016M592772); Fundamental Research Funds for the Central Universities. J.L. acknowledges the support of the following financial assistance award 70NANB18H006 from U.S. Department of Commerce, National Institute of Standards and Technology (NIST). The NIST authors acknowledge support of the NIST Quantum Based Metrology Initiative.

\section{Author contributions}
R.L. and J.L. conceived of the idea, developed the theory, and designed and initiated the experiment. R.L. carried out optical measurements and analysed the data. J.L., R.L., and R.E.L. wrote the manuscript and all authors commented on the manuscript. F.L. contributed theoretical support. R.E.L. interpreted results.  P.Z., H.G., D.P.P., and F.L. supervised the project.
\section{Methods}
\subsection{Fourier transform of LG modes}

The complex amplitude of LG modes subject to \refeq{subspace} at the beam waist is 
\begin{eqnarray}
\Psi_l(r,\phi)=\sqrt{\frac{2}{\pi l !}} \frac{1}{\sigma}(\frac{\sqrt{2}r }{\sigma})^{l} e^{-r^{2} / \sigma^{2}}e^{-i l \phi}
\label{LG_expr}
\end{eqnarray}

The Fourier transform for each term in \refeq{Inten_OAM} is
\begin{eqnarray}
\begin{split}
&P_{l_{1}, l_{2}}(r_f, \phi_f) \\
&=\mathscr{F}\left[\Psi_{l_{1}}(r, \phi) \Psi_{l_{2}}^{*}(r, \phi)\right] \\
&=(-i)^{|l_{1}-l_{2}|} \frac{\sqrt{l_{1}!l_{2}!}}{\mathrm{max}(l_{1},l_{2})!}e^{-2R^{2}} (\sqrt{2} R)^{|l_{1}-l_{2}|} \\
&\times L_{\mathrm{max}(l_{1},l_{2})}^{|l_{1}-l_{2}|}(2 R^{2})e^{i(l_{1}-l_{2})\phi_f},
\end{split}
\label{P_detailed}
\end{eqnarray}
where the following integral formulas are applied to derive the equation above,
\begin{equation}
\int_{0}^{2 \pi} e^{i \alpha \cos (\varphi-\theta)} e^{\pm i n \varphi} d \varphi=2 \pi J_{n}(\alpha) e^{i n\left(\frac{\pi}{2} \pm \theta\right)} 
\end{equation}
\begin{equation}
\begin{split}
&\int_{0}^{\infty} e^{-\alpha r^{2}} J_{n}(\beta r) r^{m+1} d r  \\
&=\frac{\beta^{n} \Gamma\left(\frac{m+n+2}{2}\right)}{2^{n+1} \alpha^{\frac{m+n+2}{2}} \Gamma(n+1)} \Pi\left(\frac{m+n+2}{2}, n+1,-\frac{\beta^{2}}{4 \alpha}\right)
\end{split}
\end{equation}
where $\alpha>0$, $\quad m+n>-1$, $J_{n}(x)$ is $n$th Bessel function, and $\Gamma(n)=(n-1)!$. $\Pi(x)$ is the confluent hypergeometric function that can be transferred to the generalized Laguerre polynomial,
\begin{equation}
L_{n}^{\alpha}(x)=e^{x} \binom{n+\alpha}{n} \Pi(\alpha+n+1, \alpha+1,-x)
\end{equation}
where $\binom{n+\alpha}{n}$ is a generalized binomial coefficient.
The orthogonal property in \refeq{Orthogonal_OAM} is guaranteed by,
\begin{equation}
\int_{0}^{\infty} e^{i m\phi} e^{i n\phi}=2 \pi\delta_{m, n}
\label{ort1}
\end{equation}

\begin{equation}
\int_{0}^{2\pi} x^{\alpha} e^{-x} L_{m}^{\alpha}(x) L_{n}^{\alpha}(x) d x=\frac{\Gamma(m+\alpha+1)}{m !} \delta_{m, n}
\label{ort2}
\end{equation}

\subsection{Experimental set-up}

The setup mainly consists of two identical arms [\reffig{setup}(a)]. In each arm, the laser beam with wavelength, $\lambda=632.8$~nm, is generated by a He-Ne laser, and mode-cleaned to TEM00 mode with single mode fibers. Then, the expanded laser beams transvers HWPs and polarization beam splitters (PBS), which in combination are used to tune the intensity of the laser beams that shine on the following SLMs.  After reflected by the SLMs, the modulated laser beams are finally combined by a beam splitter, BS3. A CCD camera placed at the Fourier plane of the SLMs records the intensity pattern of the combined beam. Computer-generated holograms (CGHs)~\cite{bolduc2013exact} are loaded on each SLM to generate arbitrary OAM near-pure states~\cite{bolduc2013exact}. Since the two He-Ne lasers are not locked, the state of the final combined laser beam is a classical mixture of two near-pure OAM states, i.e., a mixed state. The weight of each near-pure state of the final mixed state can be tuned by adjusting the angle of the HWP in each arm. The lens in front of the CCD camera is used to relocate the beam waist onto the CCD camera. It is known that different $|l\rangle$ states have different $z$-dependent Gouy phase, $\Psi(z)=(l+1)\arctan(z\lambda/\pi\sigma^2)$. By measuring at beam waist, we can eliminate the $z$-dependent Gouy phase. And the residual Gouy phase induced by the Fourier lens, $e^{-il\pi/2}$, can be cancelled by rotating the image from CCD camera by $90^{\circ}$. According to \refeq{Orthogonal_OAM}, the prior knowledge of beam waist $\sigma$ should be known to reconstruct the density matrix . To measure $\sigma$, a TEM00 mode (fundamental Gaussian mode) is generated and recorded. We fit the recorded intensity profile with a two-dimensional Gaussian function, and get the beam waist, $\sigma=0.114\pm 0.001$ mm.

\subsection{Measurement}
To generate high quality OAM state, we need to correct the aberrations in the setup in \reffig{setup}(a) by loading some correction phase pattern on the SLMs. We use Zernike polynomials to decompose the aberration in our setup. Since higher-order LG modes are more sensitive to the phase aberration, we use a LG mode with $l=50$ to measure the aberration. We first load a CGH on the phase SLM without any phase correction and fit the recorded intensity patterns from CCD with the theory intensity distribution. The fitting goodness measures the aberration. Then, we load the first 15 Zernike polynomial phase patterns one by one onto the SLMs. The magnitude of each phase pattern is adjusted to maximize the fitting goodness before proceeding to the next one. 

\begin{figure}[htb]
\centerline{\includegraphics[width=8cm]{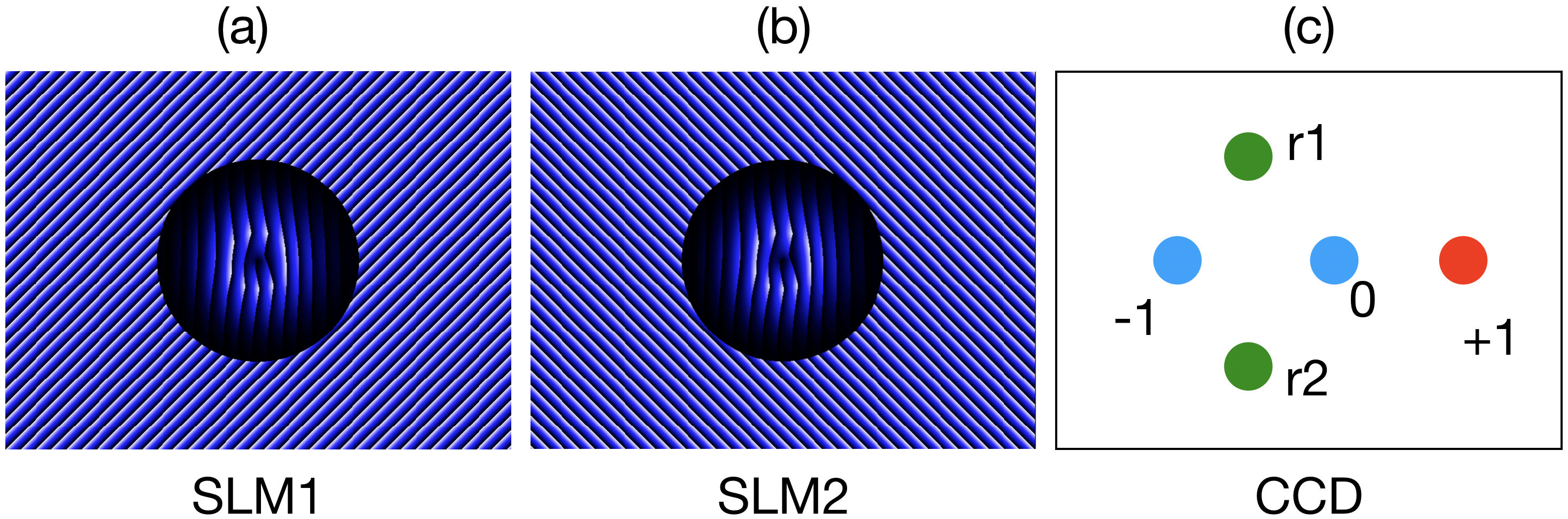}}
\caption{An example of the CGHs loaded on SLM1 (a) and SLM2 (b) for generating mixed states. (c) Recorded diffraction spots on camera.}
\label{SLM_CCD}
\end{figure}

In the experiments of generating mixed states, the weight of the two near-pure OAM states are controlled by two HWPs independently. However, the fluctuations of the power of the two independent He-Ne lasers and the coupling efficiencies of the SMFs make the weight of the two near-pure OAM states unstable. To deal with this issue, we monitor the relative power fluctuations between two arms with the same CCD camera while we are recording the mixed state intensities. We then pose-select the data with desired power ratio between the two arms. To generate the power monitoring signal, we divide each SLMs into two parts as shown in \reffig{SLM_CCD}. The circular inner parts on both SLMs load the holograms for generating near-pure OAM states and both holograms diffract incoming lasers in the horizontal direction. The outter parts load simple blazed grating to diffract incoming light at $135^{\circ}$ for SLM1 and $225^{\circ}$ for SLM2. As shown in \reffig{SLM_CCD}(c), the +1 order pattern is recorded for AHST, and the power in r1 order and r2 order are recorded for post-selecting the desired weight of the two near-pure OAM states.

\subsection{Data analysis}

In the experiment, all the intensity patterns recorded from the CCD camera have $200\times200$ pixels, which means the measurement Hilbert space has the dimension, $D=200\times200=40000$. All the OAM states to be measured live in the space spanned by $|l=0, 1, \cdots, 12\rangle$, which has the dimension, $d=13$. Thus, the discrete experimental data do satisfy tomography condiction, $D\geq d^2$. We then apply discrete Fourier transformation to the intensity patterns,
\begin{equation}
\begin{split}
&\tilde{I}(p\Delta X_f, q\Delta Y_f) =\mathscr{F}[I]\\
&=\sum_{\substack{m=-M/2+1\\n=-N/2+1}}^{M/2, N/2} I(m\Delta X, n\Delta Y) e^{-i (mp\Delta X_f \Delta X+nq\Delta Y_f \Delta Y)} 
\end{split}
\label{Discrete_Fourier}               
\end{equation}

where $M=N=200$, and $I(m\Delta X, n\Delta Y)$ is the discrete OAM intensity pattern recorded by the CCD camera. $\tilde{I}(p\Delta X_f, q\Delta Y_f)$ is discrete Fourier transformation of intensity pattern $I(m\Delta X, n\Delta Y)$, and $p, q\in\{-M/2+1, -M/2, \cdots, M/2\}$. $\Delta X$, $\Delta Y$, $\Delta X_f$ and $\Delta Y_f$ are the step sizes in the OAM intensity plane and its Fourier plane. We then substitute $\tilde{I}(p\Delta X_f, q\Delta Y_f)$ into a discrete form of \refeq{GetDensityMatrix} to calculate the density matrix,
\begin{equation}
\begin{split}
\hat{\rho}_{l_{1},l_{2}} = & \frac{C_{l_{1},l_{2}}}{A} \sum_{\substack{p=-M/2+1\\q=-N/2+1}}^{M/2, N/2} \tilde{I}(p\Delta X_f, q\Delta Y_f)P_{l_{1},l_{2}}(p\Delta X_f, q\Delta Y_f)\\
&\times
 e^{\frac{\pi^2[(p\Delta X_f)^2+(q\Delta Y_f)^2]\sigma^2}{2}}\Delta X_f\Delta Y_f,
\end{split}
\label{GetDensityMatrix_discrete}               
\end{equation}
where $P_{l_{1},l_{2}}(p\Delta X_f, q\Delta Y_f)$ is calculated using its analytical from, \refeq{P_detailed}; $A$ is a constant factor related to the total intensity. We set $A=1$ when we calculate \refeq{GetDensityMatrix_discrete} and remedy it later on by imposing the normalization of the density matrix. Note that compared to \refeq{GetDensityMatrix}, the above equation is converted from to polar coordinates to the Cartesian coordinates.
As we know, the above equation relies on the orthogonal integrals, \refeq{ort1} and \refeq{ort2}. When we perform the numberical integration in \refeq{GetDensityMatrix_discrete}, the number of discrete points will determine how precise the orthogonal integrals will be. As a result, the more pixels we use to sample the intensity, the more accurate the density matrix reconstruction will be. In our experiment, we have enough number of pixels to report more than 95\% state fidelities on all the states we generated.

\subsection{Least square method}
With the AHST protocol, a density matrix can be reconstructed simply by using \refeq{GetDensityMatrix}. However, the resulting density matrix may not be physical. In other words, there is no guarantee that the reconstructed density matrix is Hermitian and positive semi-definite. To account for this, we use Cholesky decomposition to represent an arbitrary Hermitian and positive semi-definite density matrix $\rho$ with a lower triangular matrix $T$,
\begin{equation}
\rho=\frac{T^{\dagger}T}{\text{tr}(T^{\dagger}T)}.
\label{Cholesky_decom}               
\end{equation}

For $n$-dimensional tomography space, we parametrize the density matrix $\rho$ by constructing $T$ with $n^2$ real variables $t_{i}$,
\begin{equation}
\setstackgap{L}{2\baselineskip}
\fixTABwidth{T}
    T = \scalemath{0.72}{
    \parenMatrixstack{
    t_{1} & 0 & \dots & 0 \\
    t_{n+1}+i t_{n+2} & t_{2} & \dots & 0 \\
    \vdots & \vdots & \ddots & \vdots\\
    t_{n^{2}-1}+i t_{n^{2}} & t_{n^{2}-3}+i t_{n^{2}-2}   &  \dots  & t_{n}}}
\end{equation}
We find the best estimation of $(t_{1},\cdots,t_{n^2})$ by minimizing the least square cost function,
\begin{equation}
    S=\sum_{l_{1}=0}^{n-1}\sum_{l_{2}=0}^{n-1} \left|\rho_{l_{1},l_{2}}-\rho_{l_{1},l_{2}}^{\text{oth}}\right|^2,
\end{equation}
where $\rho^{\text{oth}}$ is a density matrix that is directly calculated from \refeq{GetDensityMatrix}.

\bibliographystyle{ieeetr}

\end{document}